\documentclass[twocolumn,aps,prc,showpacs,superscriptaddress,preprintnumbers,floatfix,nofootinbib]{revtex4}
\usepackage{epsfig,graphics}
\usepackage{graphicx}
\usepackage{dcolumn}
\usepackage{bm}
\usepackage{amsmath}
\usepackage[usenames]{color}
\usepackage{ulem} 
\usepackage[colorlinks,linkcolor=blue,urlcolor=blue,citecolor=blue]{hyperref}

\begin{document}

\title{Parton collisional effect on the conversion of geometry eccentricities into momentum anisotropies in relativistic heavy-ion collisions}

\author{Long Ma}
\email[]{malong@fudan.edu.cn}
\affiliation{Key Laboratory of Nuclear Physics and Ion-beam Application (MOE), Institute of Modern Physics, Fudan University, Shanghai 200433, China}
\author{Guo-Liang Ma}
\email[]{glma@fudan.edu.cn}
\affiliation{Key Laboratory of Nuclear Physics and Ion-beam Application (MOE), Institute of Modern Physics, Fudan University, Shanghai 200433, China}
\author{Yu-Gang Ma}
\email[]{mayugang@fudan.edu.cn}
\affiliation{Key Laboratory of Nuclear Physics and Ion-beam Application (MOE), Institute of Modern Physics, Fudan University, Shanghai 200433, China}
\affiliation{Shanghai Institute of Applied Physics, Chinese Academy of Sciences, Shanghai 201800, China}


\begin{abstract}

We explore parton collisional effects on the conversion of geometry eccentricities into azimuthal anisotropies in Pb+Pb collisions at $\sqrt{s_{NN}}$ = 5.02 TeV using a multi-phase transport model. The initial eccentricity $\varepsilon_{n}$ (n = 2,3) and flow harmonics $v_{n}$ (n = 2,3) are investigated as a function of the number of parton collisions ($N_{coll}$) during the source evolution of partonic phase. It is found that partonic collisions leads to generate elliptic flow $v_{2}$ and triangular flow $v_{3}$ in Pb+Pb collisions. On the other hand, partonic collisions also result in an evolution of the eccentricity of geometry. The collisional effect on the flow conversion efficiency is therefore studied. We find that the partons with larger $N_{coll}$ show a lower flow conversion efficiency, which reflect differential behaviors with respect to $N_{coll}$. It provides an additional insight into the dynamics of the space-momentum transformation during the QGP evolution from a transport model point of view.

\end{abstract}

\pacs{25.75.-q}

\maketitle

\section{Introduction}

In high-energy heavy-ion collisions, at the extreme conditions of high temperature and high baryon density, strongly-interacting quark gluon plasma (QGP) is expected to be created. The pressure gradient of the initial compressed QGP leads to an anisotropic expansion and transfers initial-state spatial anisotropy to the final-state momentum azimuthal anisotropy, which can be measured through momentum information of the final charged hadrons~\cite{Ollitrault1992,Voloshin2008,Ackermann2001,Teaney2001,Romatschke2007,Luo2017,Chen20181}. Characterized by the flow coefficients $v_{n}$ (n = 2,3,4), azimuthal anisotropies of the final-state particles are suggested to be sensitive to not only the early stage partonic dynamics but also properties of the source~\cite{STAR2004flow,STAR2013flow,PHENIX2011flow,Margutti2019,Acharya2019}. Experimentally, systematic studies have been performed for $v_{n}$ in both large heavy-ion collision systems and small collision systems~\cite{Abelev2012,Aad2013,Chatrchyan2013,Abelev2014,Adare2015,Aidala2017,Khachatryan2015,Aidala2018,Adam2019}. Sizable $v_{n}$ observed in experiment indicates that the hot and dense QGP source is like a nearly perfect fluid.

Due to the fluid-like behavior observed for QGP, hydrodynamic models have been widely used to make predictions and are successful in describing flow harmonics at both RHIC and LHC energies~\cite{Heinz2013,Ulrich2005,Gale2013,Qiu20111441,Song2011,Song2017,Schenke2011,Alver201082}. Besides hydrodynamic models, a multiphase transport model (AMPT) is also employed in studies of anisotropic flow in high energy collisions. Including both partonic and hadronic interactions, a multiphase transport model can reasonably reproduce experimental flow measurements in both large and small collision systems~\cite{Chen2004,Bzdak2018,Nie2018,Bzdak2014,Nagle2018,Han2011,Koop2015,Huang2020}.

In recent years, an escape mechanism was proposed challenging the commonly believed hydrodynamical origin of the flow anisotropies~\cite{HeL2016,Lin2015}. It is found that instead of collectivity from partonic interactions, anisotropic parton escape dominates the flow generation in d+Au collision system as well as the semi-central Au+Au collisions. Though parton escape makes considerable contribution, it was also realized that partonic interaction is essential for generating $v_{n}$ in strong interacting systems and $v_{n}$ from partonic interaction becomes dominant in collision systems with large parton-parton interaction cross-section. Extensive studies have been performed on the harmonic flow, dihadron correlation and energy loss induced by partonic collisions~\cite{Djordjevic2006,Adil2007,Qin2008,Shin2010,Ma2014,Magl2016,Edmonds2017}. 

Theoretically, final flow anisotropy is suggested to be strongly correlated with the initial geometric anisotropy in relativistic heavy-ion collisions~\cite{Qiu201184,Lacey2014112,Alver201081,DerradideSouza2011,Alver2010104,Sorensen2007,Margutti2019,Andrade2006,Petersen2010}. It has been argued that the magnitude and trend of the partonic participant eccentricity $\varepsilon_{n}$(n = 2,3) imply specifically testable predictions for the final flow harmonics~\cite{Ma2011106,Ma2016,Wang2014b}. For a deeper understanding of the transport, it is essential to investigate the parton collisional effect on the initial geometric anisotropy as well as the conversion from coordinate space to momentum space which is expected to provide important information about the evolution dynamics of early stage .

In this paper, we present a systematic study of the partonic collision effect on the initial eccentricity and flow anisotropy in high energy Pb+Pb collisions at $\sqrt{s_{NN}}$ = 5.02 TeV, from a multi-phase transport model. Of particular interest are central collisions in which the averaged energy density is relatively higher than in non-central collisions. Furthermore, influences of partonic collision on the transfer of eccentricity anisotropy to flow anisotropy are also investigated. This paper is organized as follows: In Sec. II, a multiphase transport (AMPT) model is briefly introduced. Results and discussion are presented In Sec. III. A summary is given in Sec. IV.

\section{Model setup} 

The multi-phase transport model (AMPT) is widely used for studying transport dynamics in relativistic heavy-ion collisions. The model consists of four main components: the initial condition, partonic interaction, hadronization (quark coalescence), and hadronic interactions~\cite{Zhang2000,Lin2005}. Fluctuating initial conditions including minijet partons and soft string excitations are generated from the Heavy Ion Jet Interaction Generator (HIJING) model~\cite{Wang1991}. In the string melting scenario, both excited strings and minijet partons are melted into partons, i.e. decomposed into constituent quarks according to their flavor and spin structures. The following evolution of partonic matter is described by a parton cascade model - Zhang's parton cascade (ZPC) model~\cite{Zhang1997}, which includes elastic partonic scatterings at present.  Partons stop interacting when no parton pairs can be found within the interaction range of pQCD cross section. The transition from the partonic matter to the hadronic matter is achieved using a simple quark coalescence model which combines partons into hadrons. The final-stage hadronic interactions are modeled by a relativistic transport model (ART) including both elastic and inelastic scattering descriptions for baryon-baryon, baryon-meson and meson-meson interactions~\cite{Li1995}.

At the parton cascade stage, the differential parton-parton elastic scattering cross section is formularized based on the leading order pQCD gluon-gluon interaction:

\begin{equation}
\frac{d\sigma}{dt}=\frac{9\pi\alpha^{2}_{s}}{2}(1+\frac{\mu^{2}}{s})\frac{1}{(t-\mu^{2})^{2}},
\label{q1}
\end{equation}
where $\alpha_{s}$ is the strong coupling constant, $s$ and $t$ are the usual Mandelstam variables and $\mu$ is the Debye screening mass in partonic matter. Previous studies show that by setting proper parton scattering cross sections, AMPT model with string melting scenario has been successful in describing many experimental results in heavy-ion collisions at RHIC and LHC energies~\cite{Han2011,Ma2014,Zhou2015,Nie100519,Xuyifei2016,Wang2019,Jin2018,Xu2018}. 

In this study, we employ the string melting version of the AMPT model to focus on the partonic phase only. The parton cross section is set to be 3 mb according to Ref.~\cite{Lin2014} which reasonably reproduces the experimental results. Pb+Pb collision events are generated over a wide centrality range at center-of-mass energy of 5.02 TeV. Table.~\ref{table1} shows the definition of centrality classes and the corresponding mean number of participant nucleons. 

\begin{table}[htbp]
\caption{ Centrality classes of the AMPT events in Pb+Pb collisions at $\sqrt{s_{NN}}$ = 5.02 TeV. }
\label{table1}
\centering
\begin{tabular}{p{60pt}p{100pt}p{60pt}}
\hline
\hline
Centrality percentile & Impact parameter b(fm) &   $\left\langle{N_{part}}\right\rangle$\\
\hline
 0$\%$ - 10$\%$ & 0.0 - 4.9      & 362.8 \\
10$\%$ - 20$\%$ & 4.9 - 7.0      & 263.5 \\
20$\%$ - 30$\%$ & 7.0 - 8.6      & 188.2 \\
30$\%$ - 40$\%$ & 8.6 - 10.0     & 131.8 \\
40$\%$ - 50$\%$ & 10.0 - 11.2    & 86.1 \\
50$\%$ - 60$\%$ & 11.2 - 12.3    & 53.8 \\
\hline
\hline
\end{tabular}
\end{table}

\section{Results and Discussion}

\subsection{Parton collisions in Pb+Pb collisions}

We trace the collisional history of the initially produced partons during the source evolution in Pb+Pb collisions. The total number of parton-parton scatterings suffered by each parton before its freezing out is difined as $N_{coll}$.  

\begin{figure*}[htbp]
\centering
\resizebox{14.0cm}{!}{\includegraphics{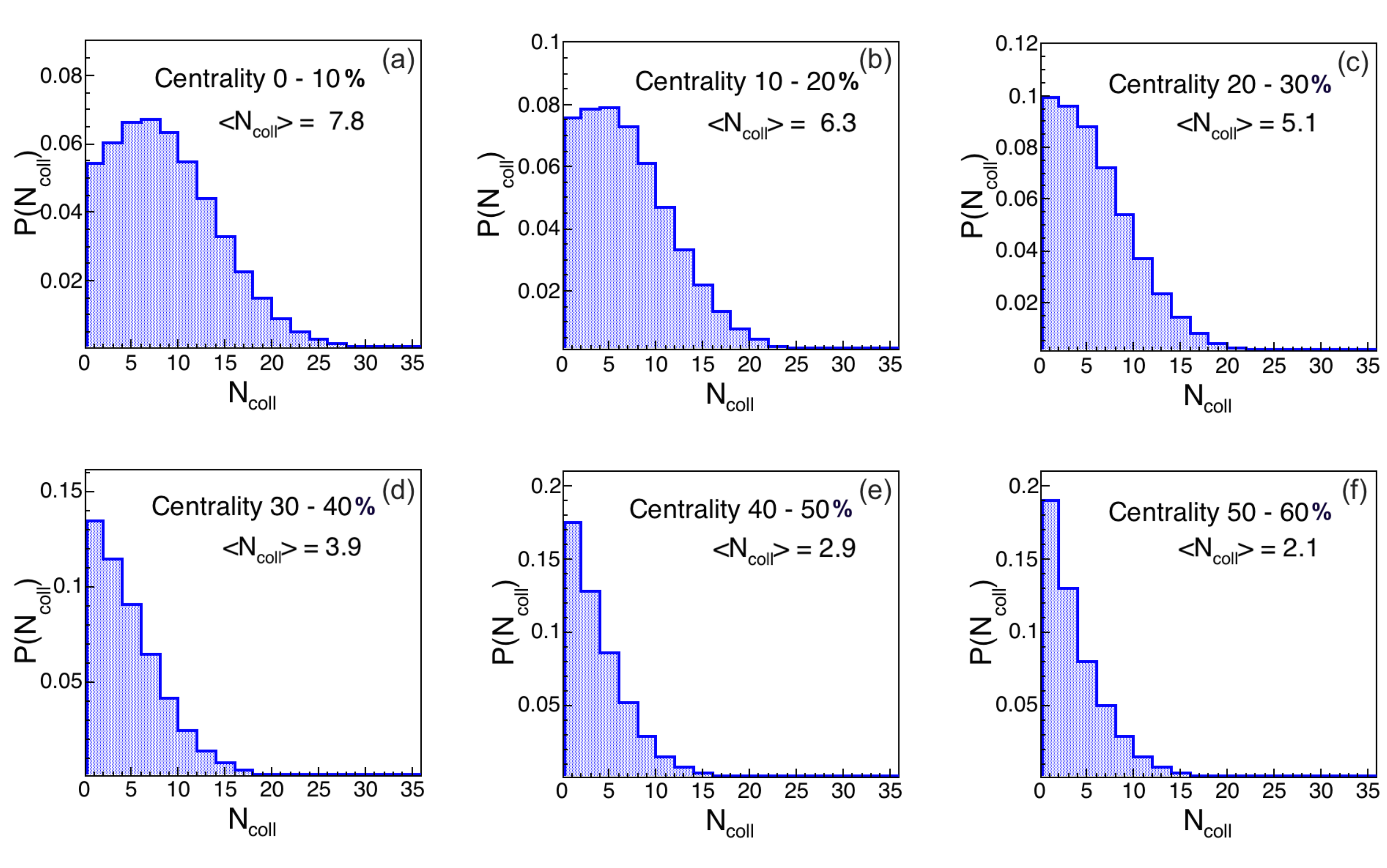}}
\caption{(Color online) Probability distributions of $N_{coll}$ for the freezeout partons for Pb+Pb collisions at $\sqrt{s_{NN}}$ = 5.02 TeV in AMPT simulation. Results are shown for six centrality classes.}
\label{f1}
\end{figure*}

Fig.~\ref{f1} shows the $N_{coll}$ distributions of the freezeout partons for different centrality classes. As expected, partons in central Pb+Pb collisions on average suffer more partonic collisions than non-central collisions before freezing out as the energy density is higher in more central collisions. The probability distribution shows a non-monotonic $N_{coll}$ dependence in central collisions which is different from that of the peripheral collisions. A peak around $N_{coll}$ $\sim$ 6 is observed for the 0-10$\%$ most central collisions. The average number of $N_{coll}$ in 0-10$\%$ centrality is found to be roughly three times as large as that in 40-60$\%$ centrality. It indicates that the fraction of partons which never collided with other partons decreases from peripheral to central collision class. 

\begin{figure*}[htbp]
\centering
\resizebox{14.0cm}{!}{\includegraphics{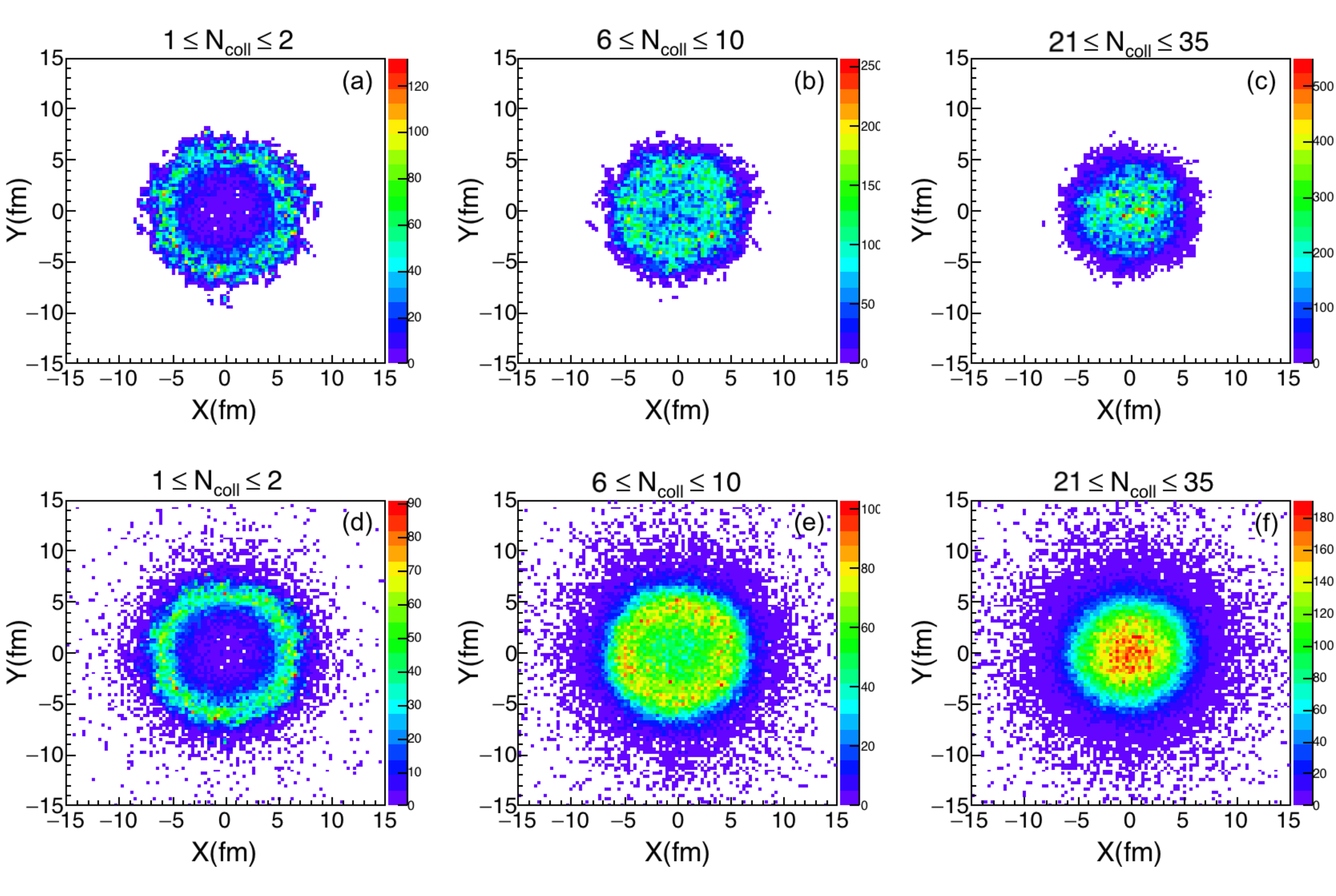}}
\caption{(Color online) Distributions of initial partons (upper panels) and freezeout partons (lower panels) for different $N_{coll}$ intervals in the transverse plane for the most central (b = 0 fm) Pb+Pb collisions at $\sqrt{s_{NN}}$ = 5.02 TeV.}
\label{f3}
\end{figure*}

\begin{figure*}[htbp]
\centering
\resizebox{14.0cm}{!}{\includegraphics{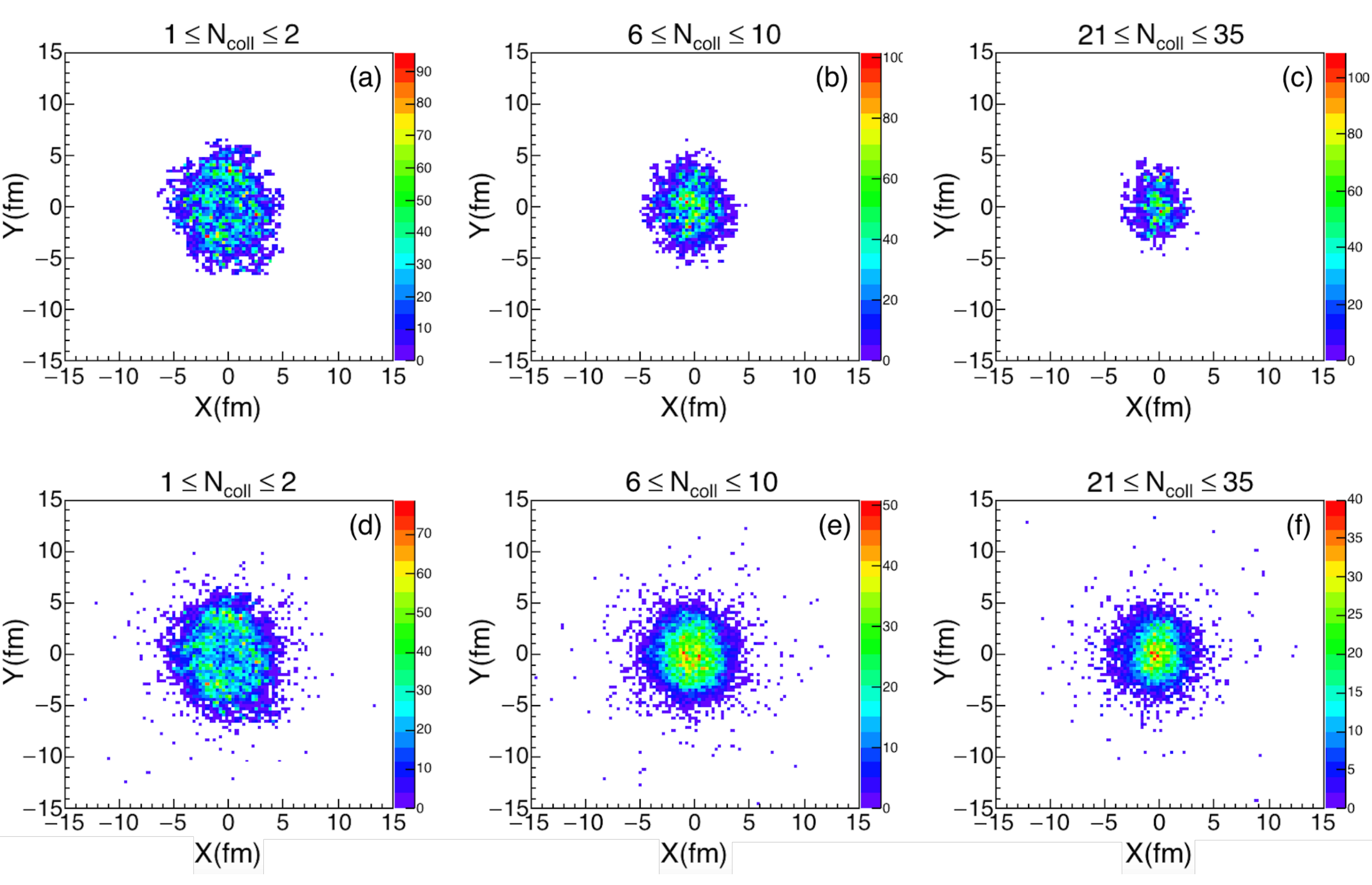}}
\caption{(Color online) Same as Fig.~\ref{f3} but for b = 11.5 fm.}
\label{f4}
\end{figure*}

We study in particular the spatial evolution of the initial partons in ultra-relativistic heavy-ion collisions. Figs.~\ref{f3} and ~\ref{f4} present the two-dimensional distributions of the initial partons [plots (a)-(c)] and freezeout partons [plots (d)-(f)] in central and peripheral Pb+Pb collisions,  where the initial parton distributions in the collision zone are compared with the final parton distributions for different $N_{coll}$ intervals. We find that partons suffering small $N_{coll}$ tend to distribute in the outer region close to the source surface whereas partons with large $N_{coll}$ are seen concentrating more in the central area. This is consistent with the expectation that due to the energy density distribution of the bulk matter, outgoing partons from the inner source suffers more collisions when passing through the bulk matter than partons close to the source surface.

\begin{figure*}[htbp]
\centering
\resizebox{16.0cm}{!}{\includegraphics{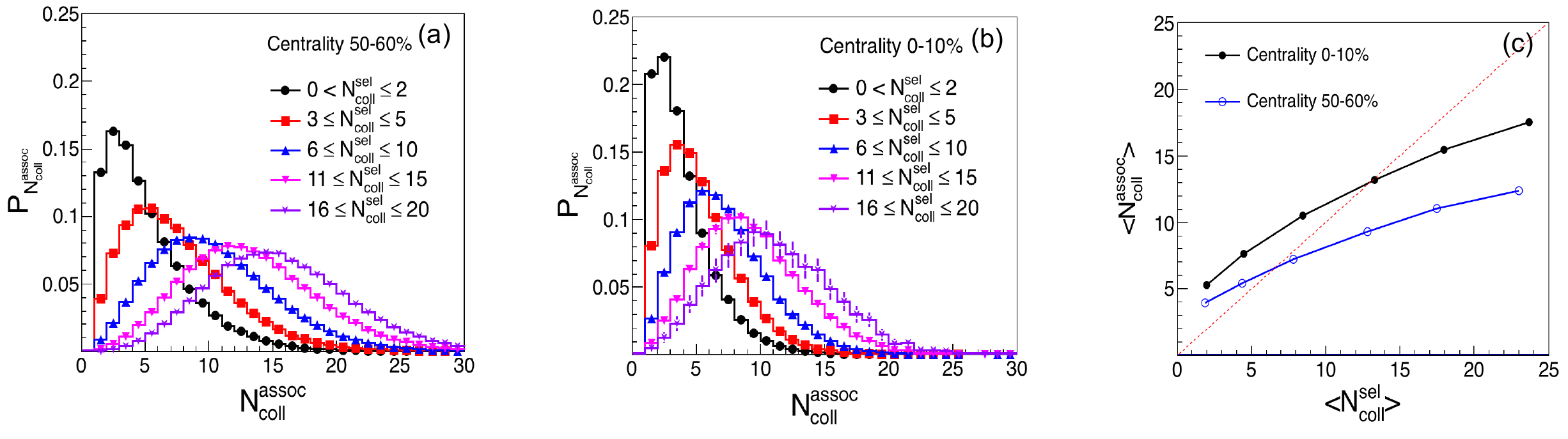}}
\caption{(Color online) The $N^{assoc}_{coll}$ probability distributions of the ``associated parton'' collided with different $N^{sel}_{coll}$ intervals of selected partons for two centrality classes of 50-60$\%$ (a)  and 0-10$\%$ (b)  in Pb+Pb collisions at $\sqrt{s_{NN}}$ = 5.02 TeV. Panel (c) shows the $\langle N^{assoc}_{coll} \rangle$ as functions of $\langle N^{sel}_{coll} \rangle$ for the two centrality classes. }
\label{f2}
\end{figure*}

We further investigate the partonic collision history of some selected partons before their freezing out. For a selected parton, we define those partons which collided with the  selected parton during its evolution as ``associated partons'' of the selected parton. The numbers of parton collisions for the selected parton and associated parton are defined as $N^{sel}_{coll}$ and $N^{assoc}_{coll}$, respectively. Fig.~\ref{f2} (a) and (b) show the $N^{assoc}_{coll}$ probability distributions for five $N^{sel}_{coll}$ intervals of the selected partons for peripheral and central Pb+Pb collisions, respectively. One can find that for a selected freezeout parton which suffered $N^{sel}_{coll}$ collisions, the number of collisions that its associated partons suffer ($N^{assoc}_{coll}$) can be distributed over a wide range. For example, for a selected parton with a small $N^{sel}_{coll}$, a long-tailed $N^{assoc}_{coll}$ distribution can be observed suggesting that many large-$N^{assoc}_{coll}$ partons play an important role in the collisional history of small-$N^{sel}_{coll}$ parton. We quantitatively extract the mean value of $N^{assoc}_{coll}$ for each $N^{sel}_{coll}$ intervals and plot the relations as shown in Fig.~\ref{f2}(c). One can find that $\langle N^{assoc}_{coll} \rangle$ is larger than $\langle N^{sel}_{coll} \rangle$ at small $\langle N^{sel}_{coll} \rangle$, which indicates the partons which suffer a small number of collisions prefer to collide with the partons which suffer a larger number of collisions.  But with the increasing of $\langle N^{sel}_{coll} \rangle$, $\langle N^{assoc}_{coll} \rangle$ tends to be close to and then less than $\langle N^{sel}_{coll} \rangle$. It indicates that the partons which suffer a large number of collisions prefer to collide with the partons which suffer a small number of collisions. In this sense, all partons are complemented with each other during the whole evolution of the partonic phase in Pb+Pb collisions.

\subsection{Collisional effect on the flow anisotropy}
 
Azimuthal anisotropy coefficients $v_{n}$ (n = 2,3..) are typically used to characterize the different orders of harmonic flow of the collision system. In simulation studies, one can calculate $v_{n}$ with respect to the participant plane of the collision event~\cite{Voloshin2007695}. The $n$th-order participant plane angle $\psi_{n}$ for a single event is in the form:

\begin{equation}
\psi_{n}\left\{PP\right\}  = \frac{1}{n}\left[ \arctan\frac{\left\langle {r^{2} \sin
(n\varphi_{PP})} \right\rangle}{\left\langle {r^{2} \cos (n\varphi_{PP})}
\right\rangle} + \pi \right],
 \label{q2}
\end{equation}
where $r$ and $\varphi_{PP}$ are the position and azimuthal angle of each parton in the transverse plane in the initial stage of AMPT and the bracket $\langle ... \rangle$ denotes per-event average. Then $v_{n}$ with respect to the participant plane $\psi_{n}\left\{PP\right\}$ is defined as

\begin{equation}
v_{n}\left\{PP\right\} = \left\langle cos[n(\phi-\psi_{n}\left\{PP\right\})] \right\rangle,
\label{q3}
\end{equation}
where $\phi$ in this study is the azimuthal angle of parton in the momentum space, and the average $\langle \cdots\rangle$ denotes event average. The above method for $v_n$ calculation is referred to as participant plane method. Participant plane method takes into account initial geometric fluctuation effect, and has been widely used in many studies~\cite{DerradideSouza2011}.

Besides the participant plane method, the multi-particle cumulant method was also proposed for studying flow via particle correlations. It has been successfully used in both model and experimental studies to quantified the harmonic flow~\cite{Bilandzic2011,Zhou2015,Abelev2014}. Usually, the two- and four-particle cumulants can be written as

\begin{equation}
C_{n}\left\{2\right\}=\langle \langle 2\rangle \rangle ,
C_{n}\left\{4\right\}=\langle \langle 4\rangle \rangle -2\langle \langle 2\rangle \rangle ^{2}.
\label{q4}
\end{equation}

The integral flow can be derived directly from two- and four-particle cumulants through the following equations

\begin{equation}
v_{n}\left\{2\right\}=\sqrt{C_{n}\left\{2\right\}},
v_{n}\left\{4\right\}=\sqrt[4]{-C_{n}\left\{4\right\}}.
\label{q5}
\end{equation}

and estimation of differential flow is according to

\begin{equation}
v^{'}_{n}\left \{ 2 \right \}=\frac{d_{n}\left \{ 2 \right \}}{\sqrt{c_{n}\left \{ 2 \right \}}},
v^{'}_{n}\left \{ 4 \right \}=\frac{d_{n}\left \{ 4 \right \}}{-c_{n}\left \{ 4 \right \}^{3/4}}
\label{q6}
\end{equation}
where the $d_{n}\left \{ 2 \right \}$ and $d_{n}\left \{ 4 \right \}$ are the two- and four-particle differential cumulants as defined in Ref.~\cite{Bilandzic2011}.

By extracting the parton information in AMPT simulation, we study the collisional effects on the development of partonic flow and eccentricity in the early stage of the heavy-ion collisions.

\begin{figure*}[htbp]
\centering
\resizebox{12.0cm}{5.0cm}{\includegraphics{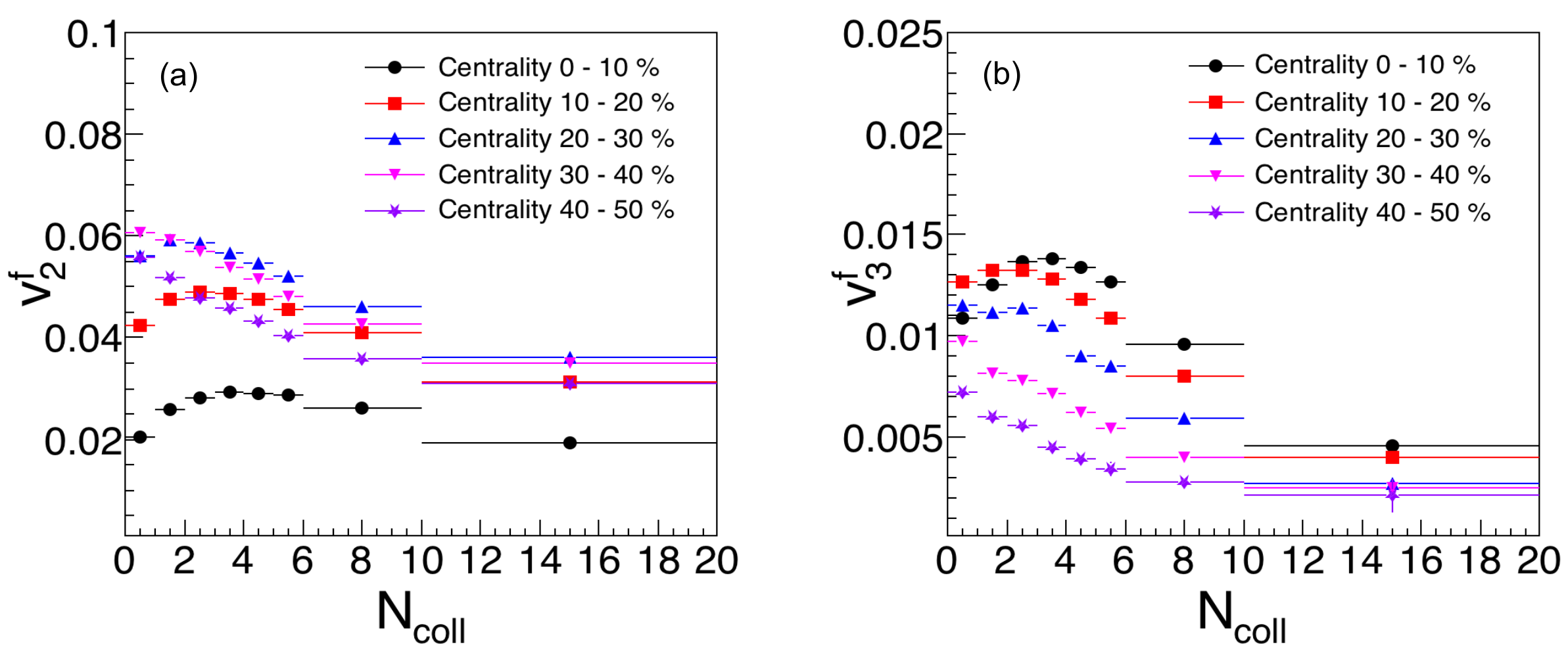}}
\caption{(Color online) $v_{n}^{f}$ (n = 2,3) of final freezeout partons from participant plane method as a function of $N_{coll}$ for Pb+Pb collisions at $\sqrt{s_{NN}}$ = 5.02 TeV in the AMPT model. Results are shown for different centality classes.} 
\label{f5}
\end{figure*}

Fig.~\ref{f5} shows the simulation results of the anisotropic flow of freezeout partons as a function of $N_{coll}$ in Pb+Pb collisions at $\sqrt{s_{NN}}$ = 5.02 TeV. The second and third order flow harmonics of the freezeout partons are defined as $v^{f}_{2}$ and $v^{f}_{3}$ respectively. Similar to the probablity distribution of $N_{coll}$, $v^{f}_{2}$ shows non-monotonic $N_{coll}$ dependence for central collisions. A maximum value of $v^{f}_{2}$ around $N_{coll}$ $\sim$ 5 is observed for 0-10$\%$ centrality. For the periphral collisions, $v^{f}_{2}$ shows a similar decreasing trend and is comparably much larger in magnitude than that of the central collisions in magnitude. It generally shows that partons with larger $N_{coll}$ tends to have smaller $v^{f}_{2}$ indicating that with increasing $N_{coll}$ the momentum azimuthal distribution of freezeout partons tend to be isotropic. This could be because large $N_{coll}$ partons come mostly from the center where the effective gradients are small. In other words, large $N_{coll}$ partons are less sensitive to the geometry than small $N_{coll}$ partons which are closer to the surface. The same conclusion also holds for $v^{f}_{3}$ but which originates basically from the initial-state fluctuations. 

\begin{figure*}[htbp]
\centering
\resizebox{12.0cm}{10.0cm}{\includegraphics{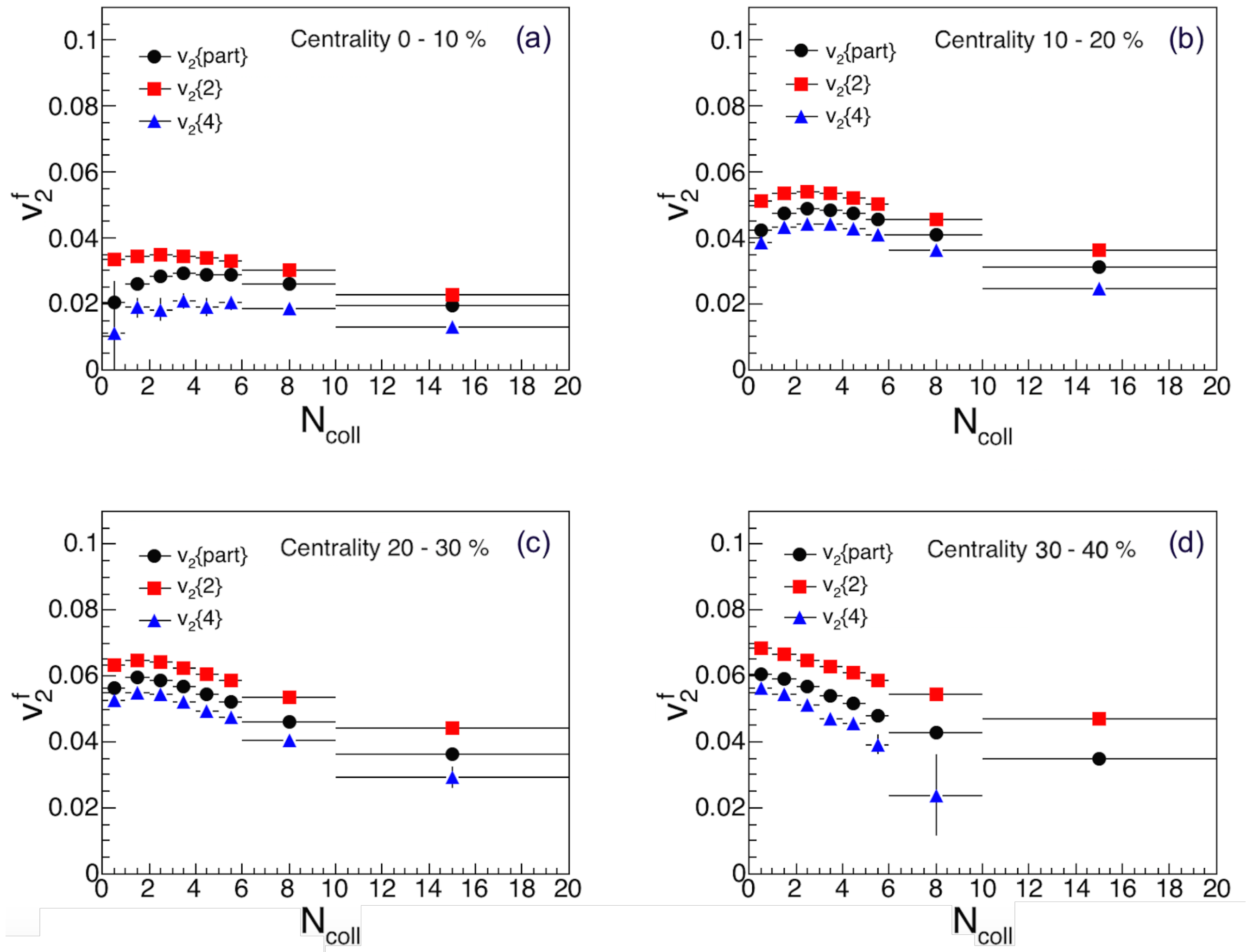}}
\caption{(Color online) $v^{f}_{2}$ of final freezeout partons from two-particle ($v_{2}\left \{ 2 \right \}$) and four-particle ($v_{2}\left \{ 4 \right \}$) cumulant method as a function of $N_{coll}$ for Pb+Pb collisions at $\sqrt{s_{NN}}$ = 5.02 TeV in the AMPT model. Comparisons are made with participant plane method for different centrality classes.}
\label{f6}
\end{figure*}

Besides the participant plane method, we further studied $v^{f}_{n}$ based on multi-particle cumulant methods. Fig.~\ref{f6} shows the two-particle ($v_{2}\left \{ 2 \right \}$) and four-particle ($v_{2}\left \{ 4 \right \}$) cumulant flow results. Comparisons are made with the results from the participant plane method. It is generally found that $v^{f}_{2}$ from cumulant methods are similar in trend with the $v^{f}_{2}$ results from participant plane method. An ordering of $v_{2}\left \{ 2 \right \} > v_{2}\left \{ part \right \} > v_{2}\left \{ 4 \right \}$ is observed, because $v_{2}\left \{ 2 \right \}$ involves flow fluctuations but $v_{2}\left \{ 4 \right \}$ suppresses non-flow contributions.

\begin{figure*}[htbp]
\centering
\resizebox{14.0cm}{8.5cm}{\includegraphics{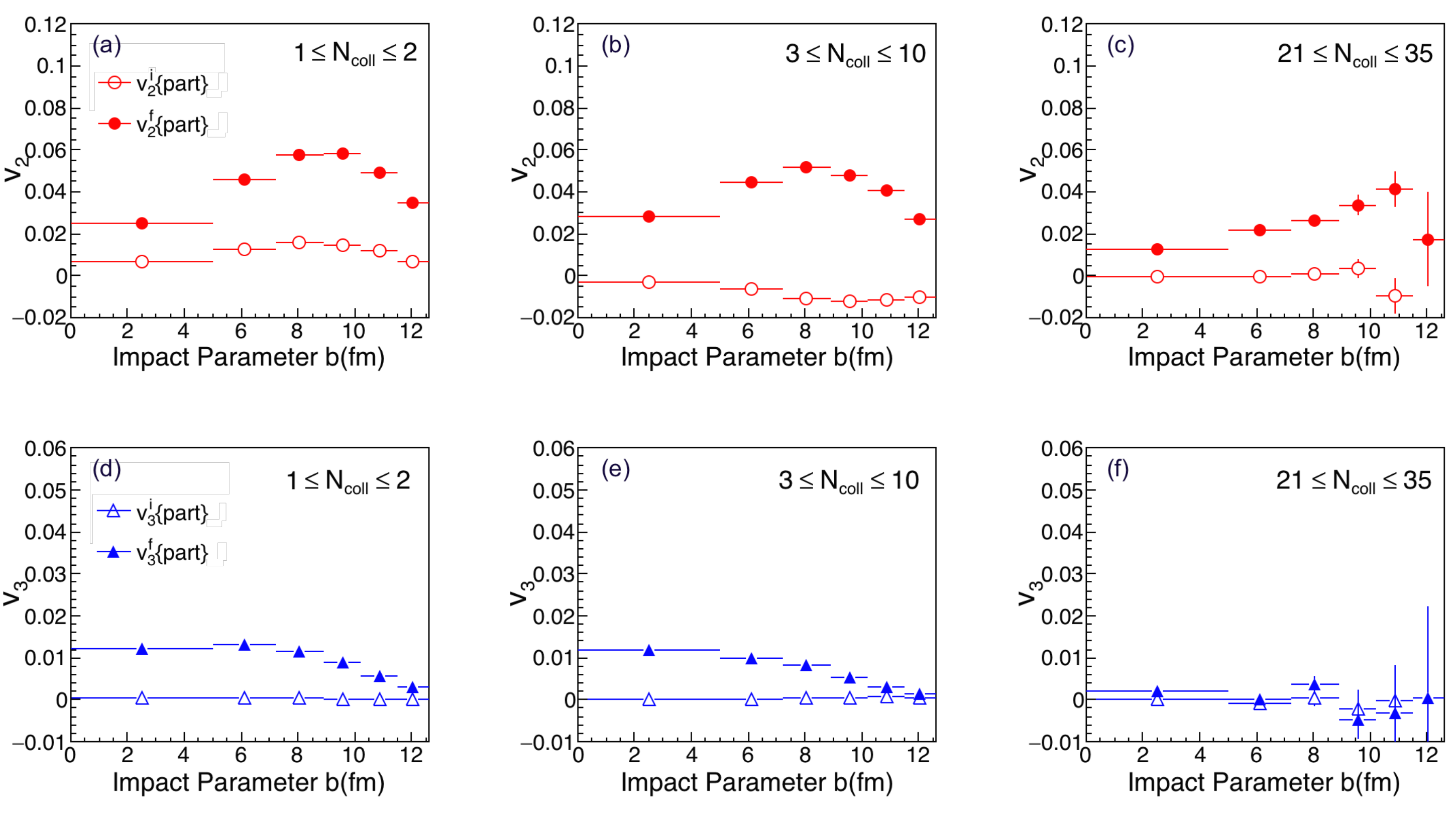}}
\caption{(Color online) $v^{i}_{n}$ of initial state partons and $v^{f}_{n}$ of final freezeout partons as a function of impact parameter for different $N_{coll}$ intervals.}
\label{f7}
\end{figure*}

Towards a more quantitative study, we compare the flow anisotropies of the initial partons ($v^{i}_{n}$) and final freezeout partons ($v^{f}_{n}$) for n = 2,3. Note that the averaged $v_{n}$ for all the initial partons is zero due to the isotropy of initial azimuthal distribution in the AMPT model. As can be seen in Fig.~\ref{f7}, showing the change of $v_{n}$ from the initial to the final stage of the partonic evolution, the parton-parton collisions generally make $v_{2}$ and $v_{3}$ increase for different $N_{coll}$ regions. For the partons with a larger number of $N_{coll}$, the change of $v_{2}$ and $v_{3}$ is smaller, since they are more probably located at the center of the source where more collisions may randomize their motion. 

\begin{figure*}[htbp]
\centering
\resizebox{12.0cm}{10.0cm}{\includegraphics{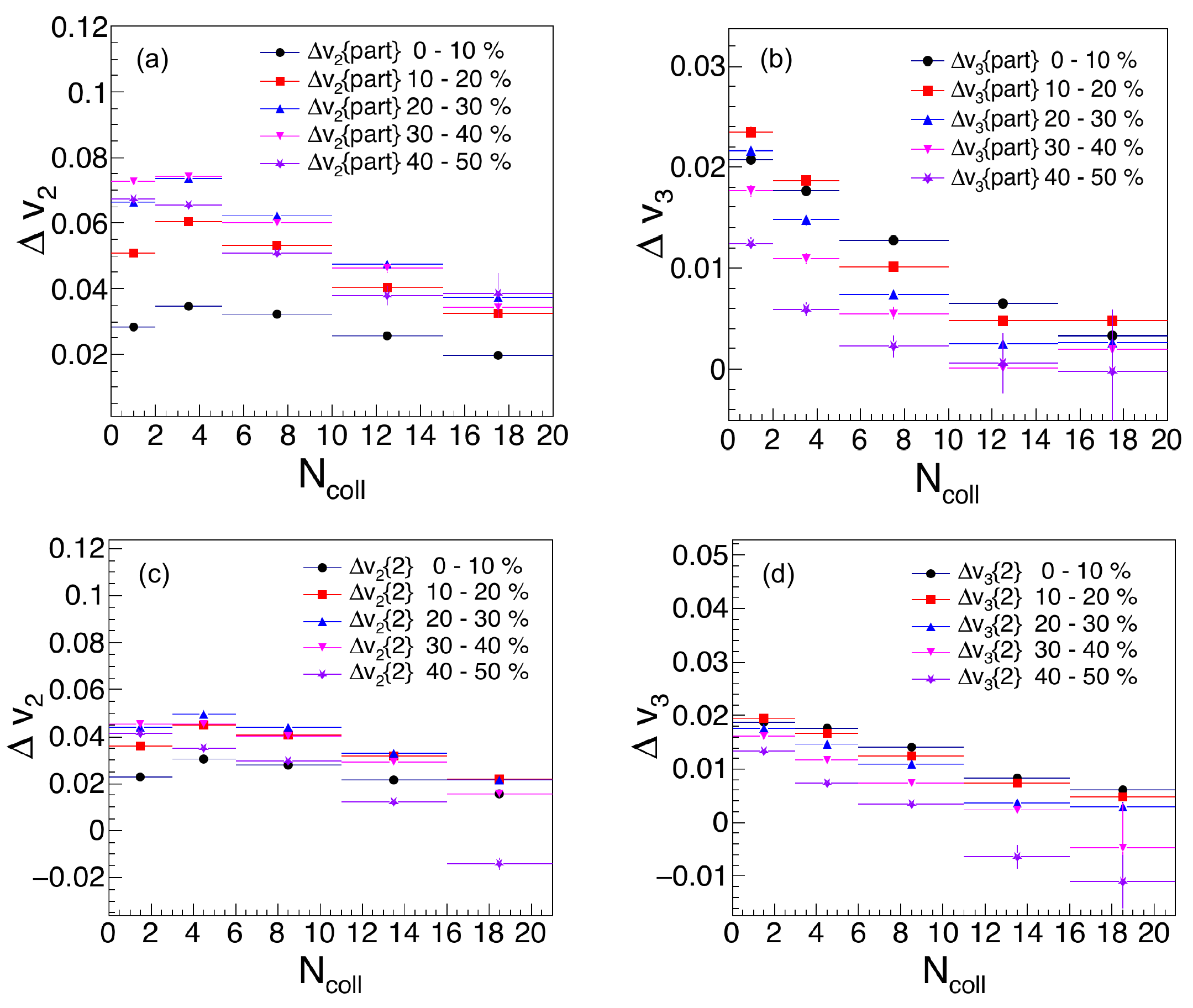}}
\caption{(Color online) The AMPT results of $\Delta$$v_{n} =v^{f}_{n}-v^{i}_{n}$ (n = 2,3) as a function of $N_{coll}$, where the upper panels show $\Delta$$v_{n}$ from participant plane method and the lower panels show $\Delta$$v_{n}$ from two-particle cumulant methods.}
\label{f8}
\end{figure*}

In order to quantitatively study the collisional effect on the flow harmonics, we examine the change of parton $v_{n}$ after $N_{coll}$ collisions, i.e. $\Delta$$v_{n}$ = $v^{f}_{n}$ - $v^{i}_{n}$. Fig.~\ref{f8} shows the $\Delta$$v_{n}$ (n = 2,3) in Pb+Pb collisions as a function of $N_{coll}$. Results are compared for different flow methods. Significant centrality dependence can be seen for $\Delta$$v_{n}$. It is interesting to see that in central collisions $\Delta$$v_{2}$ shows non-monotonic $N_{coll}$ dependence whereas $\Delta$$v_{3}$ presents monotonic $N_{coll}$ dependence. As shown in Fig.~\ref{f7}, the initial intrinsic parton $v_{n}$ is quite tiny, the gain in $v_{n}$ is primarily due to the partonic scatterings throughout the source evolution. In general, $\Delta$$v_{n}$ decreases with increasing of $N_{coll}$ indicating that small $N_{coll}$ partons contribute to most of the flow anisotropies.

\subsection{Collisional effect on the initial eccentricity}

Initial geometry anisotropy of the QGP matter is a main source responsible for generating the final flow anisotropy in relativistic heavy-ion collisions. Thus, it is important to study the partonic effect on the initial spatial anisotropy. In nuclear-nuclear collisions, the spatial anisotropy of the collision zone in the transverse plane (perpendicular to the beam direction) can be characterized by the eccentricity $\varepsilon _{n}$. It has been argued that the magnitude and trend of the eccentricity imply testable predictions for final-state hadronic flow~\cite{Miller2003,Lacey2010,Drescher2007,Broniowski2007}. 

The definition of the $n$th-order harmonic eccentricity in the coordinate space of the participant nucleons or partons for single collision event is in the form:

\begin{equation}
\varepsilon _{n}\left\{part\right\}  = \frac{{\sqrt {\left\langle {r^{n} \cos (n\varphi)} \right\rangle ^2 + \left\langle {r^{n} \sin (n\varphi)}
\right\rangle ^2 } }}{{\left\langle {r^{n} } \right\rangle }},
\label{q7}
\end{equation}
where $r$ and $\varphi$ are position and azimuthal angle of each nucleon or parton in the transverse plane. $\varepsilon _{n}\left\{part\right\}$ characterizes the eccentricity through the distribution of participant nucleons or partons which naturally contains event-by-event fluctuation. $\varepsilon _{n}\left\{part\right\}$ defined in this way is usually named as ``participant eccentricity''.  

\begin{figure*}[htbp]
\centering
\resizebox{14.0cm}{8.5cm}{\includegraphics{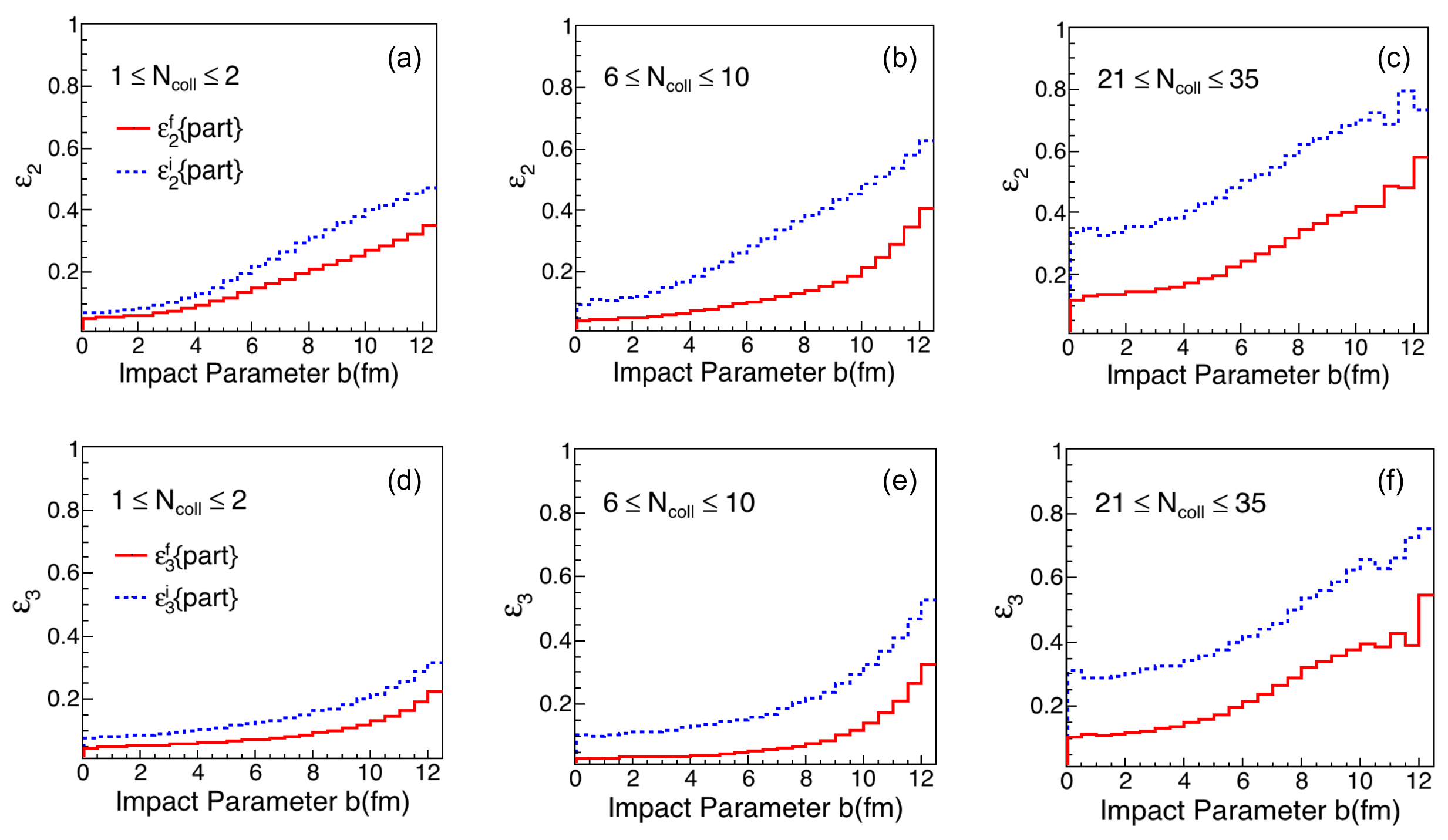}}
\caption{(Color online) $\varepsilon_{2}^{i,f}\left\{part\right\}$ (upper panels) and  $\varepsilon_{3}^{i,f}\left\{part\right\}$ (lower panels) of the initial state and final freezeout partons as a function of impact parameter for Pb+Pb collisions at $\sqrt{S_{NN}}$ = 5.02 TeV. Simulation results are shown for different $N_{coll}$ intervals. }
\label{f9}
\end{figure*}

We study the parton collisional effects on the partonic eccentricity in Pb+Pb collisions. Fig.~\ref{f9} shows the AMPT results of the second and third order eccentricities calculated with Eq.~(\ref{q7}). Eccentricities of the initial and final freezeout partons are denoted as $\varepsilon^{i}_{n}\left\{part\right\}$ and $\varepsilon^{f}_{n}\left\{part\right\}$ respectively. Similarly to the flow harmonics, partonic scattering is found to play an important role in the evolution of eccentricities. $\varepsilon_{n}$ of the freezeout partons is larger at larger $N_{coll}$. One can see in the figures that parton collisions generally reduce $\varepsilon_{n}$ which is consistent with our expectation - during the expansion of the QGP source, the transition of the initial pressure gradient from coordinate space to the momentum space will significantly diminish the spatial anisotropy.

\begin{figure*}[htbp]
\centering
\resizebox{12.0cm}{5.0cm}{\includegraphics{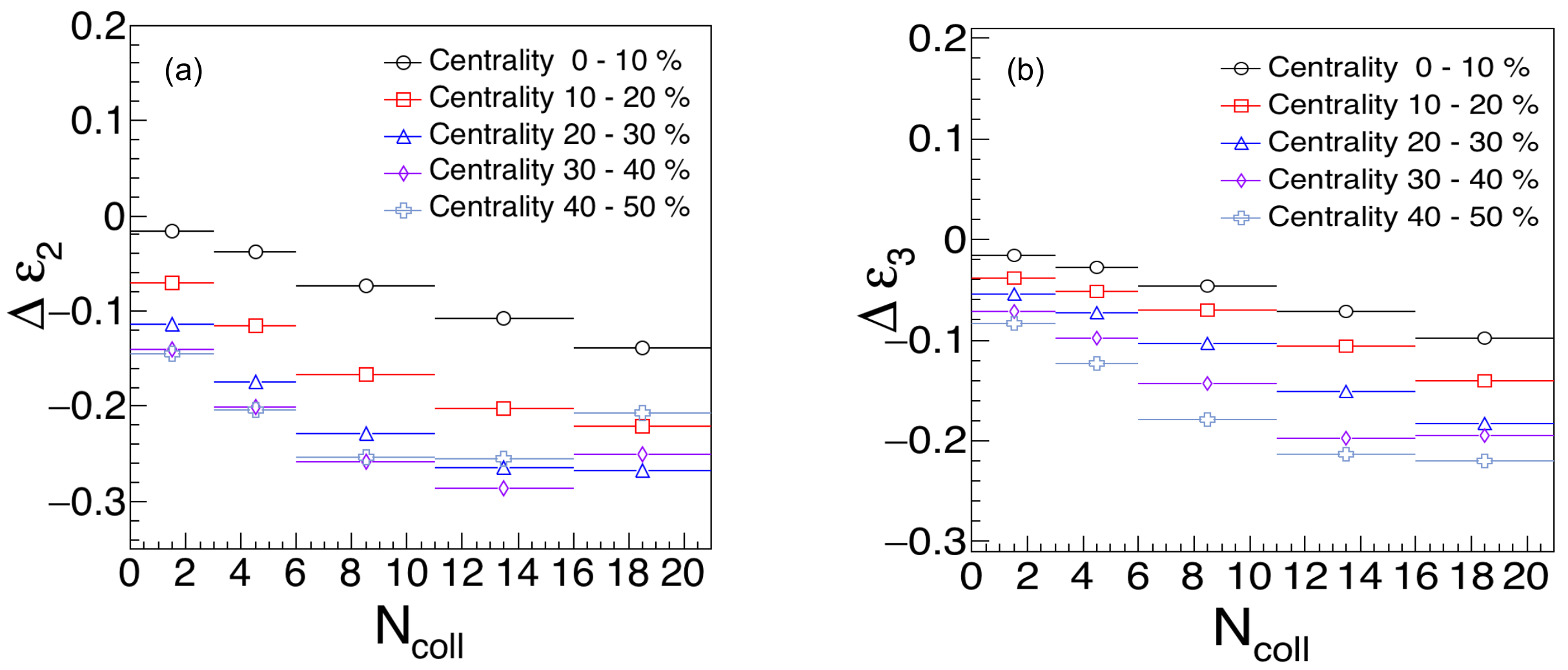}}
\caption{(Color online) The AMPT results of $\Delta$$\varepsilon_{n}$=$\varepsilon^{f}_{n}\left\{part\right\}$ - $\varepsilon^{i}_{n}\left\{part\right\}$ (n=2,3) as a function of $N_{coll}$ for different centrality classes.}
\label{f10}
\end{figure*}

In addition, we studied $\Delta$$\varepsilon_{n}$=$\varepsilon^{f}_{n}\left\{part\right\}$ - $\varepsilon^{i}_{n}\left\{part\right\}$ as a function of the number of parton collisions for different centrality classes. The results for second and third order harmonics are shown in Fig.~\ref{f10} (a) and (b), respectively. We find that $\Delta$$\varepsilon_{2}$ and $\Delta$$\varepsilon_{3}$ exhibit clear decreasing $N_{coll}$ dependences. Quantitative difference are seen between the results for $\Delta$$\varepsilon_{2}$ and $\Delta$$\varepsilon_{3}$ , because $\varepsilon_{3}$ is purely driven by initial fluctuations but $\varepsilon_{2}$ is driven by initial geometry.

\subsection{Collisional effect on the flow response to the eccentricity}

Impressive progress has been made in studying the final-state flow response to the initial eccentricity in relativistic heavy-ion collisions~\cite{Petersen2012,Niemi2013}. The success of hydrodynamical models tells us that elliptic flow $v_2$ and triangular flow $v_3$ are mainly driven by the linear response to the initial ellipticity and triangularity of the source geometry. As space-momentum correlation is also expected to be built during the partonic evolution stage, quantitative study of the partonic flow response in a event-by-event transport model is also important for understanding the development of final flow . 

\begin{figure*}[htbp]
\centering
\resizebox{12.0cm}{5.0cm}{\includegraphics{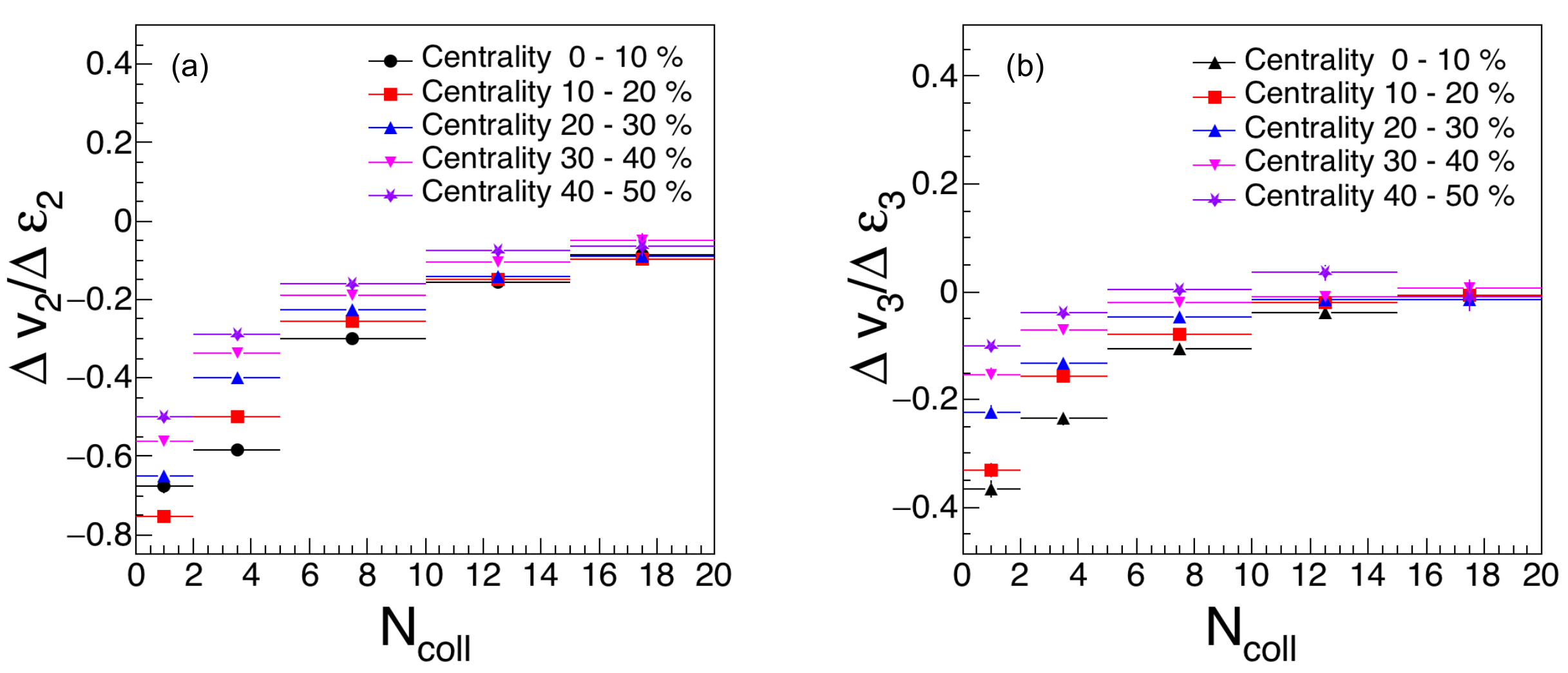}}
\caption{(Color online) Conversion efficiency $\Delta$$v_{n}$/$\Delta$$\varepsilon_{n}$ for n=2 (left panel) and n=3 (right panel) as a function of $N_{coll}$ for Pb+Pb collisions at 5.02 TeV. }
\label{f11}
\end{figure*}

\begin{figure*}[htbp]
\centering
\resizebox{12.0cm}{5.0cm}{\includegraphics{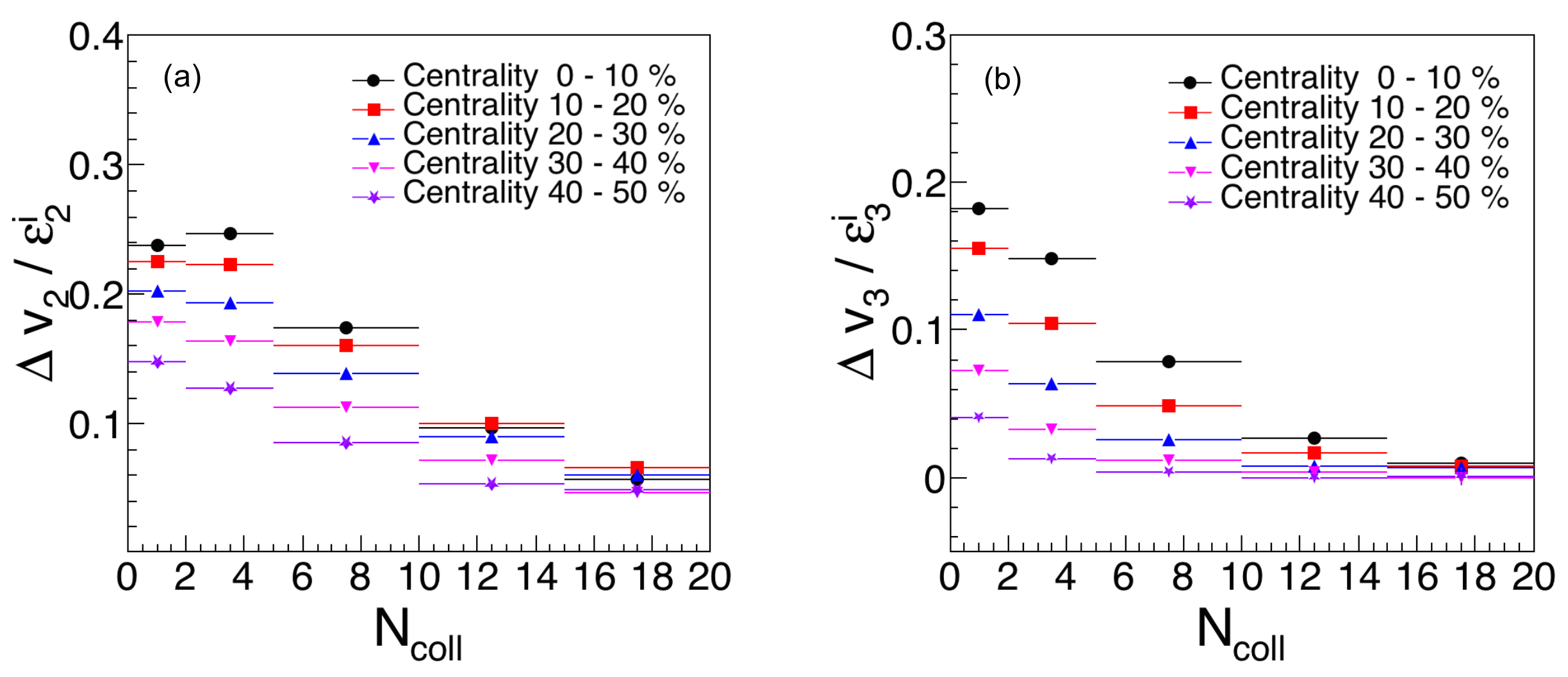}}
\caption{(Color online) $\Delta$$v_{n}$/$\varepsilon^{i}_{n}$ for n=2 (left panel) and n=3 (right panel) as a function of $N_{coll}$ for Pb+Pb collisions at 5.02 TeV, where $\varepsilon^{i}_{n}$ is the initial partonic eccentricity.}
\label{f12}
\end{figure*}

Based on the AMPT model simulations, we studied $N_{coll}$ effects on the flow response by looking into the ratio $\Delta$$v_{n}$/$\Delta$$\varepsilon_{n}$. Since $\Delta$$\varepsilon_{n}$ (n = 2,3) are negative and $\Delta$$v_{n}$ (n = 2,3) are positive over all the $N_{coll}$ classes, one could take the absolute value of $\Delta$$v_{n}$/$\Delta$$\varepsilon_{n}$ as an estimation of the flow conversion efficiency. Fig.~\ref{f11} and ~\ref{f12} show the results for the flow conversion efficiency with respect to $\Delta$$\varepsilon_{n}$ and $\varepsilon^{i}_{n}$ as a function of $N_{coll}$. Considering absolute value, results in both figures show similar trend. The ratio of $\Delta$$v_{n}$/$\Delta$$\varepsilon_{n}$ presents obvious $N_{coll}$ dependences for different centrality classes. We observe that for both elliptic and triangular flow the conversion efficiency is strongest in the collision class of 0-10\%. It indicates that more collisions in more central collisions help transfer $\varepsilon_{n}$ into $v_n$, which is a normal concept about the flow conversion efficiency which is an integral effect of all $N_{coll}$ partons. For the differential $N_{coll}$ dependence, $\Delta$$v_{n}$/$\Delta$$\varepsilon_{n}$ (n = 2,3) presents a smooth increasing trend from small to large $N_{coll}$, which indicates that the larger $N_{coll}$ is, the lower the flow conversion efficiency is. The feature seems against common sense, but it can be understood through the above results that parton collisional contribution to flow change $\Delta$$v_{n}$ is more significant at smaller $N_{coll}$ whereas that to eccentricity change $\Delta$$\varepsilon_{n}$ is more significant at larger $N_{coll}$, i.e. changes of $\Delta$$v_{n}$ and $\Delta$$\varepsilon_{n}$ are not in sync with respect to $N_{coll}$. But since small-$N_{coll}$ and large-$N_{coll}$ partons are complemented with each other during the evolution, it is actually hard to fairly say which should be given the first credit to the generation of the final flow. In the limit of long evolution time, final eccentricity $\varepsilon^{f}_{n}$ is supposed to approach zero, and we observe the similar results for $\Delta$$v_{n}$/$\varepsilon^{i}_{n}$ except with the opposite sign, as shown in Fig. ~\ref{f12}.

\section{Summary}

In summary, we studied initial partonic flow anisotropy ($v_{n}$) and spatial anisotropy ($\varepsilon_{n}$) in Pb+Pb collisions at center-of-mass energy of 5.02 TeV using a multi-phase transport model. By tracing the partonic cascade history in AMPT, the effect of the parton-parton collisions was intensively investigated. We find that partonic collision plays an important role in the development of flow anisotropies in heavy-ion collisions. We find that the partons which suffer a small number of collisions prefer to collide with the partons which suffer a larger number of collisions, and vice versa. The change of $v_{n}$ decreases with increasing of $N_{coll}$ indicating that small $N_{coll}$ partons contribute to most of the flow anisotropies. However, the change of eccentricity is more significant for the large-$N_{coll}$ partons. As a result, the partons with larger $N_{coll}$ show a lower flow conversion efficiency, which reflect the differential behaviors of the flow conversion efficiency with respect to $N_{coll}$. However, since small-$N_{coll}$ and large-$N_{coll}$ partons are always complemented with each other, it is hard to rank their roles in generating flow. 

However, one has to be aware that although the AMPT model provides an effective tool to simulate and study parton-parton collisions in relativistic heavy-ion collisions, the initial partonic source configured using constituent quarks in the string-melting scenario could introduce some intrinsic bias into our study, since the created QGP should consist of both current quarks and gluons. In addition,  the approximation of the model treatment of the parton interactions is in a way analogous to gluon-gluon elastic interaction based on the leading order pQCD cross section which could also introduce some bias and lead to an incomplete or improper description, since the QGP evolution involves non-perturbative QCD processes. Nevertheless, such a simplified picture of the partonic evolution in this model is expected to provide some guides to the study of the effect on the conversion rules of the initial eccentricity to the final flow anisotropy.

As anisotropic flow of initial partons will transfer to the final hadrons which are formed from the coalescence of freeze-out quarks in the transport model, further study by tracing the hadronization and hadronic evolution of particles would be important for fully understanding the complete behavior of the anisotropic flow. We postpone such investigations for our future study.

\section*{Acknowledgements}

We thank A. Bzdak and Z. -W. Lin for helpful discussions. This work is supported in part by the National Natural Science Foundation of China under Contracts No. 11961131011, No. 11890710, No.11890714, No.11835002, No.11421505, No.11905034, the Key Research Program of Frontier Sciences of Chinese Academy of Sciences under Grant No. QYZDJ-SSW-SLH002, the Strategic Priority Research Program of Chinese Academy of Sciences under Grant No. XDB34030000,  and the Guangdong Major Project of Basic and Applied Basic Research No. 2020B0301030008.

\bibliographystyle{apsrev.bst}


\end{document}